\documentclass[ reprint,showpacs,
 superscriptaddress,
 pra,
]{revtex4-2}
\usepackage{graphicx} 
\usepackage{amsfonts}
\usepackage{mathrsfs,amssymb}
\usepackage[version=4]{mhchem}
\usepackage[dvipsnames]{xcolor}
\usepackage{mathtools}
\usepackage[titletoc]{appendix}
\usepackage{dcolumn} 

\usepackage[usenames,dvipsnames]{xcolor} 
\usepackage{soul}                        
\usepackage[normalem]{ulem}              

\newcommand{\g}{_\text{g}}
\newcommand{\e}{_\text{e}}
\newcommand{\D}{_\text{D}}
\newcommand{\Z}{_\text{Z}}

\newcommand{\ket}[1]{|#1\rangle} 

\renewcommand\vec{\mathbf}

\makeatletter
\newcommand{\newparallel}{\mathrel{\mathpalette\new@parallel\relax}}
\newcommand{\new@parallel}[2]{%
  \begingroup
  \sbox\z@{$#1T$}
  \resizebox{!}{\ht\z@}{\raisebox{\depth}{$\m@th#1/\mkern-5mu/$}}%
  \endgroup
}
\makeatother

\allowdisplaybreaks

\graphicspath{ {./figures/} }

\begin{document}

\title{On Demand magnetic-Doppler nuclear frequency comb memory for hard X-ray photons}

\author{Yanli Shi}
\affiliation{Institute for Quantum Science and Engineering, Department of Physics and
  Astronomy, Texas A\&M University, College Station, Texas 77843, USA}
\author{Xiwen Zhang}%
 \email{xiwen@tamu.edu}
\affiliation{Institute for Quantum Science and Engineering, Department of Physics and
  Astronomy, Texas A\&M University, College Station, Texas 77843, USA}
\author{Yuri Shvyd’ko}
\affiliation{Advanced Photon Source, Argonne National Laboratory, Argonne, Illinois 60439, USA}
\affiliation{Institute for Quantum Science and Engineering, Department of Physics and
  Astronomy, Texas A\&M University, College Station, Texas 77843, USA}
\author{Olga Kocharovskaya}
\email{kochar@physics.tamu.edu}
\affiliation{Institute for Quantum Science and Engineering, Department of Physics and
  Astronomy, Texas A\&M University, College Station, Texas 77843, USA}

\date{\today}

\begin{abstract}
Nuclear quantum memories in the hard X-ray regime offer some key advantages over their optical counterparts, such as broader bandwidth and lower background noise. A Doppler frequency comb protocol has been theoretically proposed [X. Zhang \textit{et al.}, Phys. Rev. Lett. \textbf{123}, 250504 (2019)] and recently demonstrated experimentally [S. Velten \textit{et al.}, Sci. Adv. \textbf{10}, eadn9825 (2024)] for the storage and retrieval of X-ray photons. However, achieving on-demand retrieval remains challenging because of the requirement for precise and synchronous mechanical motion of multiple absorbers.
We propose a hybrid, magnetic-Doppler nuclear frequency comb composed of Doppler-shifted resonant absorbers with lifted nuclear spin degeneracy, which expands the Doppler comb structure. By synchronously reversing the directions of both the magnetic fields and absorber velocities, the system achieves time-reversed phase evolution dynamics that allows for efficient on-demand photon retrieval with significantly reduced mechanical complexity.
\end{abstract}


\maketitle

\section{Introduction}
\label{sec_introduction}

Quantum computing and quantum communication have rapidly evolved in the era of quantum technology, driving breakthroughs in the fields of fast problem solving, cyber-security, quantum chemistry, and beyond. 
A key requirement for these advances is the development of efficient quantum memories (QMs) that can temporarily store quantum bits and retrieve them with high precision on demand.
Photons with well-defined temporal waveforms are naturally suited for reliable and fast transmission of qubits, but their flying nature poses a challenge for long-term storage.

To overcome this, a variety of QM schemes have been developed in the optical frequency regime, using electromagnetically induced transparency (EIT), off-resonant Raman scattering, engineered inhomogeneous broadening in gradient echo memory (GEM) as well as atomic frequency comb (AFC) protocols~\cite{Lei23Hosseini}.
These approaches have been implemented on a wide range of material platforms, such as rare-earth-ion-doped crystals, color centers, and alkali vapors.
Meanwhile, at hard X-ray frequencies, photon qubits possess inherent advantages of larger bandwidth, smaller diffraction limit, greater penetration depth, and substantially lower detection noise --- making their storage both compelling and essential for quantum information applications.
However, a straightforward extension of the traditional QM protocols developed in the optical range proves to be difficult.
The EIT, off-resonant Raman, and AFC schemes rely on strong coherent driving that is unavailable in the hard X-ray range, while implementing a GEM scheme remains challenging because of the need for a large external gradient magnetic field.

With very few exceptions, nuclear transitions lie in the hard X-ray regime and beyond, making them a natural platform for hosting hard X-ray wave packets. In solid-state materials, the M{\"o}ssbauer effect enables resonant, recoilless scattering of X-ray photons~\cite{Wertheim64}. Unlike electronic transitions in atomic ensembles, M{\"o}ssbauer solids offer compelling advantages by providing a unique combination of significantly higher atomic number density and longer coherence time even at room temperature. Combined with the short wavelength of hard X-ray photons from radioactive sources, the X-ray -- nuclear quantum interface can provide a compact and robust platform for quantum information applications.

The extremely sharp M{\"o}ssbauer resonances enable high-fidelity, phase-coherent X-ray photons manipulation that is, in several respects, comparable --- or even superior --- to optical quantum control. Quantum control of single X-ray photons and coherent manipulation
in nuclear resonant scattering were established in the 20th century through numerous M{\"o}ssbauer source– and synchrotron-based experiments. These demonstrated coherent control of X-ray photon states using M{\"o}ssbauer nuclei, including mechanical phase modulation~\cite{Helisto91Katila, Shvydko92Smirnov}, dynamic magnetic switching~\cite{Shvydko93Smirnov, Shvydko94Gerdau, Shvydko96Schindelmann, Shvydko95Ruter}, radio-frequency techniques~\cite{Lippmaa95Katila}, and related time-domain methods~\cite{Shvydko91Hertrich}. 
In recent years, coherent control of nuclear quantum states and hard X-ray photons has been further explored,
including mechanical oscillation~\cite{Vagizov14Kocharovskaya, Shakhmuratov15Kocharovskaya, Radeonychev20Kocharovskaya} and sudden displacement~\cite{Shakhmuratov11Kocharovskaya, Shakhmuratov13Kocharovskaya, Heeg21Evers} of M{\"o}ssbauer absorbers, 
parametric resonance between nuclear vibration and Bragg modes~\cite{Zhang13Svidzinsky}, X-ray cavities~\cite{Haber16Rohlsberger, Haber17Rohlsberger}, magnetically controlled nuclear excitations~\cite{Liao12Keitel, Wang18Liao}, 
parametric down conversion~\cite{Adams00Novikov, Shwartz12Harris}, second harmonic generations~\cite{Shwartz14Harris}, etc.
Alongside these, numerous other seminal works in the field of X-ray quantum optics laid the foundation for modern X-ray quantum information technologies (see Refs.~\cite{Shvydko89Smirnov, Smirnov99Smirnov, Kagan99Kagan, Shvydko99,  Kuznetsova17Kocharovskaya} and references therein).

Recently, the Doppler nuclear frequency comb (D-NFC, or simply DFC) protocol --- based on multiple M{\"o}ssbauer absorbers moving synchronously at different, uniformly spaced velocities to form a comb-like absorption structure --- was proposed for the storage of single hard X-ray photons~\cite{Zhang19Kocharovskaya}, and was subsequently demonstrated experimentally in a set of $^{57}$Fe M{\"o}ssbauer foils using synchrotron radiation and X-ray cavity ~\cite{Velten24Rohlsberger}.
But the retrieval time in the DFC protocol is pre-determined by the chosen velocity spacing. On demand retrieval could be achieved by rapidly and simultaneously reversing all absorber velocities~\cite{Zhang19Kocharovskaya}. However, its experimental demonstration presents a significant technological challenge, particularly as the number of absorbers increases.

In this work, we show that the number of absorbers can be reduced  --- without significantly compromising the performance of QM --- by exploiting magnetic hyperfine levels allowing simultaneously for realization of on demand retrival through simultaneous reversal of both absorber velocities and the internal magnetic field. 
 The use of the hyperfine structure to reduce the number of moving absorbers in the DFC protocol was previously proposed~\cite{Yeh19Liao}. However, the possibility of on-demand retrieval through simultaneous reversal of both absorber velocities and the internal magnetic field has not been discussed so far.
The latter is motivated by the earlier experimental demonstration~\cite{Shvydko95Ruter} of time-reversed quantum beats pattern in nuclear Bragg scattering of synchrotron radiation, achieved by reversing the strong hyperfine magnetic field in iron borate (\ce{$^{57}$FeBO3}).
Yet, in a single absorber such reversed M{\"o}ssbauer time spectra do not  constitute a viable QM. 
In strong (internal) magnetic fields, the hyperfine transitions of large-spin nuclei typically yield unequally spaced spectral lines because of  pronounced electric quadrupole interactions, while nuclei with small spin ($I=1/2$) provide too few spectral lines.
In both cases, premature re-emission of hard X-ray photons prior to magnetic field reversal could strongly undermine the storage efficiency and fidelity.
By combining the DFC configuration with magnetic hyperfine levels, however, an evenly spaced nuclear frequency comb (NFC) can be achieved, enabling on-demand retrieval of stored hard X-ray photons with reduced mechanical complexity through simultaneously reversing the absorber velocities and the internal magnetic field.

Specifically, we demonstrate an on-demand QM for single hard X-ray photons using such a hybrid magnetic hyperfine-Doppler NFC, implemented with only three \ce{^{57}FeBO3} targets --- two moving at equal and opposite velocities relative to a stationary one.
Mechanical synchronization is readily achieved with two identical piezoelectric actuators driven in opposite phases, while a strong internal hyperfine field can be reversed by applying a weak external magnetic field.



\section{Magnetic-Doppler nuclear frequency comb}
\label{sec_DZNFC}

The technological challenge of Doppler-only NFC, D-NFC, lies in the large number of synchronously moving M{\"o}ssbauer absorbers required, as each absorber contributes only one single comb tooth. 
This difficulty increases dramatically for on-demand retrieval, which requires rapid and simultaneous reversal of the velocities of all targets.
A straightforward way to reduce the number of absorbers is by exploiting the internal degrees of freedom of the nuclear transition. In principle, both magnetic hyperfine and Zeeman sublevels could be utilized. 
However, in \ce{^{57}Fe}, producing an adjancent differential Zeeman splitting, or NFC comb spacing, of just two natural linewdiths would require an external magnetic field of $\approx 1$~T. 
Neither fast reversal of such a strong external field within $\sim 10$~ns, far shorter than the nuclear lifetime, appears feasible for on-demand QM retrieval.
In contrast, the $14.4$~keV \ce{^{57}Fe} nuclear transition in iron borate crystal naturally exhibits lifted degeneracy due to the strong internal hyperfine magnetic field of $B_\text{hf} \gtrsim 30$~T present in the (0001) easy-magnetization plane of the hexagonal setting of its rhombohedral structure, corresponding to a comb spacing of $60$ natural linewidths, and the direction of this internal hyperfine field can be reversed on the time scale $\lesssim 4$~ns~\cite{Shvydko94Gerdau}.

\begin{figure}[hbt!]
    \centering
    \includegraphics[width=\linewidth]{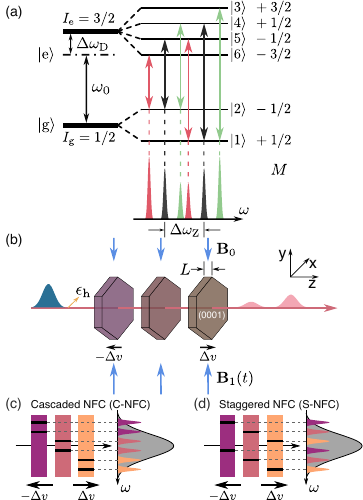}
    \caption{
    Illustration of the magnetic-Doppler hybrid NFC.
    (a)
    The energy-level diagram of the $\hbar\omega_0=14.4$~keV nuclear transition in $\ce{^{57}FeBO3}$. The ground state $\ket{\text{g}}$ and excited state $\ket{\text{e}}$ split into 2 and 4 sublevels, respectively, labeled according to their nuclear magnetic quantum numbers $M$. The 2 transitions used in this work satisfying the selection rule $M\g-M\e=0$ are spectrally separated by $\Delta \omega_Z$.
    (b)
    Three thin $\ce{^{57}FeBO3}$ M\"ossbauer absorbers, each of thickness $L$, resonantly interact with a linearly polarized hard X-ray photon, where the magnetic polarization vector $\epsilon_\text{h}$ is along $x$-direction. These absorbers move along $z$-axis with constant velocity spacing $\Delta v$, resulting in Doppler detunings $\Delta \omega_D$ on the transition frequencies shown in (a).
    A uniform external magnetic field $\vec{B}_\text{ex}=\vec{B}_0+\vec{B}_1(t)$, where $\vec{B}_0$ and $\vec{B}_1(t)$ are a constant and a time-dependent component, respectively, is applied along the $y$-direction to all absorbers to align the internal hyperfine magnetic fields $\vec{B}_\text{hf}$ to the $x$-axis.
    On-demand quantum memory of incident hard X-ray photons is realized by abruptly changing $\mathbf{B}_1$ and $\Delta \vec{v}$ at time $T_\text{sw}$ to simultaneously reverse the directions of the total external field $\vec{B}_\text{ex} \to -\vec{B}_\text{ex}$ and the moving velocity spacing $\Delta \vec{v} \to -\Delta \vec{v}$.
    Panels (c, d) illustrate two spectral-spacial configurations of the nuclear frequency combs, (c) the cascaded NFC (C-NFC) and (d) the staggered NFC (S-NFC), respectively.
    }
    \label{fig:FeBO3_level_comb}
\end{figure}

The ground state $\ket{I\g=1/2}$ and the excited state $\ket{I\e=3/2}$ of \ce{^{57}FeBO3} possess nuclear magnetic dipole moments $\mu\g=0.0906 \mu_\text{N}$ and $\mu\e = -0.1549 \mu_\text{N}$, corresponding to gyromagnetic ratios $\gamma\g = 8.68 \times 10^6$~rad~s\textsuperscript{-1}~T\textsuperscript{-1} and $\gamma\e = -4.95 \times 10^6$~rad~s\textsuperscript{-1}~T\textsuperscript{-1}, respectively~\cite{Stone14Stone}.
Here $\gamma = \mu / (I \hbar)$, $I$ is the nuclear spin quantum number, $\mu_\text{N}$ is the nuclear magneton, $\hbar = h/(2\pi)$ is the reduced Planck constant, and the subscripts “g” and “e” denote ground and excited states, respectively.
The excited state also carries an electric quadrupole moment $Q\e = 0.16$~barn~\cite{Stone14Stone}, leading to a quadrupolar interaction with the electric field gradient oriented along the crystallographic axis [0001].
At room temperature, the quadrupole coupling constant $C_q = e^2 q Q/h = -8.71$~MHz, where $-e$ is the electron charge and $-eq$ is the electric field gradient along [0001], is one-third of the excited-state Larmor frequency $\omega_{\text{L}e} = \gamma\e B_\text{hf}$~\cite{Lyubutin22Alekseeva}. As a result, the quadrupole interaction can be neglected at temperatures well below the N\'{e}el point of $348$~K, yielding the energy-level diagram shown in Fig.~\ref{fig:FeBO3_level_comb}~(a).


\ce{^{57}FeBO3} is an antiferromagnet exhibiting weak ferromagnetism. A small external magnetic field, $B_\text{ex} \lesssim 100$~G, applied in the (0001) plane is sufficient to magnetize the crystal such that $\vec{B}_\text{hf}$ nearly perpendicular to $\vec{B}_\text{ex}$.
For simplicity, we consider an incident resonant photon that is linearly polarized, with its magnetic field unit vector $\vec{\epsilon}_\text{h}$ aligned along the $x$-axis, as illustrated in Fig.~\ref{fig:FeBO3_level_comb}~(b).
When $\vec{B}_\text{hf} \perp \vec{\epsilon}_\text{h}$, the \ce{^{57}Fe} nuclei are excited by the hard X-ray photon under the selection rule $\Delta M = M\e - M\g = \pm 1$, resulting in four spectral lines in the absorption spectrum, where $M$ denotes the nuclear magnetic quantum number. However, these four lines do not form a frequency comb structure, as they are not equally spaced.
In contrast, when $\vec{B}_\text{hf} \newparallel \vec{\epsilon}_\text{h}$ the absorption spectrum consists of only two lines corresponding to transitions with $\Delta M = 0$, which likewise do not form a comb.
These two lines have equal intensities and are the focus of the present work.

The limited number of spectral teeth originates from the intrinsically low spin quantum number of the nuclear ground state. To increase the number of teeth, multiple M{\"o}ssbauer absorbers with equally spaced central transition frequencies can be employed using the D-NFC technique. The resulting comb incorporates both Doppler and hyperfine frequency components, forming a magnetic hyperfine-Doppler hybrid NFC shown in Fig.~\ref{fig:FeBO3_level_comb}.

As illustrated in Fig.~\ref{fig:FeBO3_level_comb}~(b), $N$ identical \ce{^{57}FeBO3} resonant absorbers, each with thickness $L$ and \ce{^{57}Fe} number density $\mathcal{N}$, are placed sequentially along the X-ray propagation direction $z$, which is orthogonal to their (0001) basal planes. In this work, we consider the case of $N=3$ to minimize experimental complexity.
The $n$-th absorber moves along $z$ with a velocity given by $v_n = (n - 2)\Delta v \left[1 - 2 \vartheta\D \Theta(t - T_\text{sw})\right]$, where $n = 1, 2, 3$, $\Delta v$ is the velocity spacing between neighboring absorbers, $T_\text{sw}$ denotes the time at which the velocities and/or magnetic field are switched, $\vartheta\D = 0$ or $1$, and $\Theta$ is the Heaviside step function.
A D-NFC with angular frequency spacing
\begin{align}
  \Delta \omega\D = \frac{\Delta v}{c} \omega_0
\end{align}
is thereby formed due to Doppler effect, where $c$ is the speed of light in vacuum and $\omega_0$ is the angular transition frequency of the $14.4$~keV nuclear resonance.
A uniform, weak external magnetic field $B_\text{ex} = B_0\left[1 - 2\vartheta\Z \Theta(t - T_\text{sw})\right]$ is applied along the $y$-axis, consisting of a static component $B_0 \sim 10$~G~\cite{Shvydko94Gerdau} and a time-dependent part $B_1 = -2\vartheta\Z B_0 \Theta(t - T_\text{sw})$, with $\vartheta\Z = 0$ or $1$. The purpose of $B_\text{ex}$ is to magnetize \ce{^{57}FeBO3} absorbers, aligning their internal hyperfine fields $\vec{B}_\text{hf}$ to the magnetic polarization $\vec{\epsilon}_\text{h}$ of the incident X-ray, oriented along the $x$-axis. 
The strong hyperfine magnetic field $\vec{B}_\text{hf}$ produces multiple resonant absorption peaks, with angular frequency spacing
\begin{align}
  \Delta\omega\Z=(\gamma_\text{g} - \gamma_\text{e})B_\text{hf}
\end{align}
according to the $\Delta M = 0$ selection rule [see Fig.~\ref{fig:FeBO3_level_comb}~(a)],
where $\gamma_\text{g}$ and $\gamma_\text{e}$ are gyromagnetic ratios of the ground and excited states, respectively.
Subscripts  ``D" and ``Z" denote Doppler and hyperfine or Zeeman components, respectively.

The binary parameters $\vartheta\D$ and $\vartheta\Z$ determine whether switching is applied: $\vartheta\D = \vartheta\Z = 0$ corresponds to static NFCs, leading to pseudo on-demand quantum memory with a fixed storage time, while $\vartheta\D = \vartheta\Z = 1$ enables full on-demand operation by reversing both velocities and hyperfine field directions. A partial switching case with $\vartheta\D \neq \vartheta\Z$ is discussed in the Supplementary Information~\cite{SI}.

Compared to the original D-NFC approach, the number of M{\"o}ssbauer absorbers is reduced by a factor of non-degenerate transitions $K$. For \ce{^{57}FeBO3}, $K = 2\min \{I_g, I_e\}+1 = 2$, corresponding to the smaller multiplicity of the ground and excited states, while for non-magnetic system such as \ce{^{57}Fe}-enriched stainless steel, $K=1$ as $\Delta \omega\Z = 0$.
Despite the very much limited $K$ in iron borate, it still permits the implementation of a 6-tooth NFC using just three absorbers.
The mechanical control of these three absorbers requires almost no synchronization, because the two outer absorbers can be driven by a single voltage signal with opposite phases, while the central target remains static throughout, as illustrated in Fig.~\ref{fig:FeBO3_level_comb}~(b).
As shown in Sections \ref{sec_static_NFC} and \ref{sec_on_demand_QM}, the resulting NFC, composed of six spectral components with frequency spacing $\Delta \omega$, is sufficient to demonstrate high-fidelity storage of a single X-ray wave packet with a relatively simple temporal profile, such as a Gaussian.

Depending on the spatial distribution of these spectral components, the magnetic hyperfine-Doppler NFC can be configured in two distinct ways, as illustrated in Fig.~\ref{fig:FeBO3_level_comb}~(c,~d).
In the cascaded configuration (C-NFC), shown in Fig.~\ref{fig:FeBO3_level_comb}~(c), all hyperfine components from each absorber remain spectrally packed, and absorbers are arranged sequentially along the beam path according to their resonant frequencies. This yields a comb with frequency spacing $\Delta \omega = \Delta \omega\Z$, while the Doppler frequency spacing is $\Delta \omega\D = K \Delta \omega\Z$.
In contrast, in the staggered configuration (S-NFC), shown in Fig.~\ref{fig:FeBO3_level_comb}~(d), hyperfine components of the same order from different absorbers are spectrally grouped together, effectively interleaving the Doppler-shifted components across the spectrum. This results in a comb with frequency spacing $\Delta \omega = \Delta \omega\D$, where the hyperfine splitting satisfies $\Delta \omega\Z = N \Delta \omega\D$.
Throughout this this work, we use $K=2$ and $N=3$ for both C-NFC and S-NFC configurations. For comparison, we also consider Doppler-only NFC (D-NFC) with $K=1$ and both $N=3$ and $N=6$.

Since the amplitude of the hyperfine magnetic field in \ce{^{57}FeBO3} can only be tuned within a limited range of $\sim 15$ -- $55$~T via slow temperature variation~\cite{Lyubutin22Alekseeva}, the two magnetic-Doppler NFC configurations enable adjustment of the comb teeth spacing from $\Delta \omega = \Delta \omega\Z$ down to $\Delta \omega\Z / N$ by mechanical motion.
This, in turn, allows for fast and robust tuning of the NFC echo period $T_0 = 2\pi / \Delta \omega$ as well as the total comb bandwidth $NK\Delta \omega$.
As $2T_0$ sets the upper limit of the storage time for the discussed on-demand QM scheme (see later discussion) and $2\pi / (NK\Delta \omega)$ sets the lower limit of the duration of the stored photon, these two configurations --- as well as hybrid configurations intermediate between them when more absorbers are used --- offer a flexible and versatile platform for NFC-based quantum memory.


The proposed magnetic-Doppler NFC is described by the Maxwell-Bloch equations~\cite{SI} (see also Ref.~\cite{Shvydko99} for other approaches)
\begin{align} 
& \left(\frac{\partial}{\partial z}+\frac{1}{c}\frac{\partial}{\partial t}\right)\mathcal{B}^{(n)}\left(t,z\right)=\sum_{M_{e}M_{g}}\mathcal{M}_{M_{e}M_{g}}^{\prime(n)}\left(t,z\right), \label{eq:MB_BM3_1}\\
& \frac{\partial}{\partial t}\mathcal{M}_{M_{e}M_{g}}^{\prime(n)}\left(t,z\right)=\left(-\frac{\Gamma}{2}+i\Delta_{M_{e}M_{g}}^{(n)}\right)\mathcal{M}_{M_{e}M_{g}}^{\prime(n)}\left(t,z\right) \notag \\
& \qquad \qquad \qquad \qquad \quad - \frac{\Gamma\xi_{M_{e}M_{g}}^{(n)}}{4L}\mathcal{B}^{(n)}\left(t,z\right), \label{eq:MB_BM3_2}
\end{align}
where $\mathcal{B}$ is the slowly varying amplitude of the magnetic field component of the hard X-ray photon, $\mathcal{M}^\prime$ is proportional to the slowly varying amplitude of the induced magnetization wave in the nuclear medium, the superscript ``$(n)$'' denotes quantities for the $n$-th absorber ($n=1,2,\cdots, N$), 
$\Gamma/(2\pi) = 1.13$~MHz is the spontaneous decay rate of the excited state, and $\Delta_{M_{e}M_{g}}^{(n)}$ and $\xi_{M_{e}M_{g}}^{(n)}$ are the one-photon detuning and partial optical thickness, respectively,
for the transition between states $\left|M_{e}\right>$ and $\left|M_{g}\right>$
of the $n$-th absorber. The one-photon detuning arises from both Doppler and Zeeman shifts, $\Delta_{M_{e}M_{g}}^{(n)} = -\left[ n-\left(N+1\right)/2\right] \Delta \omega_D - M_< \Delta \omega_Z$, where $M_<$ is the magnetic quantum number of $I_<$, with $I_< = \min \left\{I_\text{g}, I_\text{e}\right\}$.
The photon satisfies the continuity condition across absorbers such that the output from the $n$-th absorber serves as the input to the $(n+1)$-th absorber.
To assess the intrinsic quantum memory performance of the magnetic-Doppler NFC, we neglect off-resonant losses due to photon–electron scatterings.
These losses can be straightforwardly accounted for by including an overall attenuation factor $\exp(- \mathcal{N} \sigma_\text{ph} NL)$ in the storage efficiency, where $\sigma_\text{ph}$ is the photoelectric scattering cross section.

\section{Hard X-ray quantum memory by static nuclear frequency comb}
\label{sec_static_NFC}

Hard X-ray quantum memory can be directly realized using the proposed magnetic-Doppler NFC without any velocity or magnetic field switching, i.e., with $\vartheta\D = \vartheta\Z = 0$. In this static setup, constructive intra- and inter-absorber interference of polarization waves at discrete times $t = p T_0$ leads to the formation of hard X-ray photon echoes. Here, $p = 1, 2, \cdots$ denotes the echo order, and $T_0 = 2\pi / \Delta \omega$ is the rephasing period determined by the frequency spacing $\Delta \omega$ of the comb.
Unless otherwise specified, we focus exclusively on the first echo ($p = 1$), which corresponds to the earliest and typically most efficient retrieval of the incident photon. This constitutes a pseudo on-demand QM protocol with a fixed storage time $T_0$ set by the spectral structure of the comb.

An NFC can be characterized by the number of teeth $NK$, the comb finesse $\mathcal{F}=\Delta\omega /\Gamma$, and the total resonant optical thickness $\xi$.
To facilitate a meaningful comparison between the magnetic-Doppler NFC and the Doppler-only NFC (D-NFC), we introduce the individual optical thickness, defined as the optical thickness per absorber per spectral tooth: $\xi^0 = \xi/(NK)$. 
The effective optical thickness is given by $\xi_\text{eff} = \xi/\mathcal{F}$, and analogously, the individual effective optical thickness is $\xi^0_\text{eff} = \xi^0/\mathcal{F}$.
Similar to the D-NFC protocol, when the comb finesse and bandwidth are sufficiently large, the first-echo retrieval in the magnetic-Doppler NFC is~\cite{SI}
\begin{align}
    \mathcal{B}_\text{out}(t) = \frac{\pi \xi_\text{eff}^0} {2} \exp\left(- \frac{\pi \xi_\text{eff}^0} {4}\right) \exp\left(- \frac{\pi} {\mathcal{F}}\right) \mathcal{B}_\text{in}(t-T_0),
\end{align}
where $\mathcal{B}_\text{in}(t)$ and $\mathcal{B}_\text{out}(t)$ are the slowly-varying amplitude of the incident and output photons, respectively.
The efficiency (see the Supplementary Information~\cite{SI} for the definition) of the first echo can be optimized under $\xi_\text{eff}^0 = 4 / \pi$, achieving a storage efficiency of up to $\eta = 54\%$ in the limit of varnishing decoherence. 

\begin{figure*}[hbt!]
    \centering
    \includegraphics[width=\linewidth]{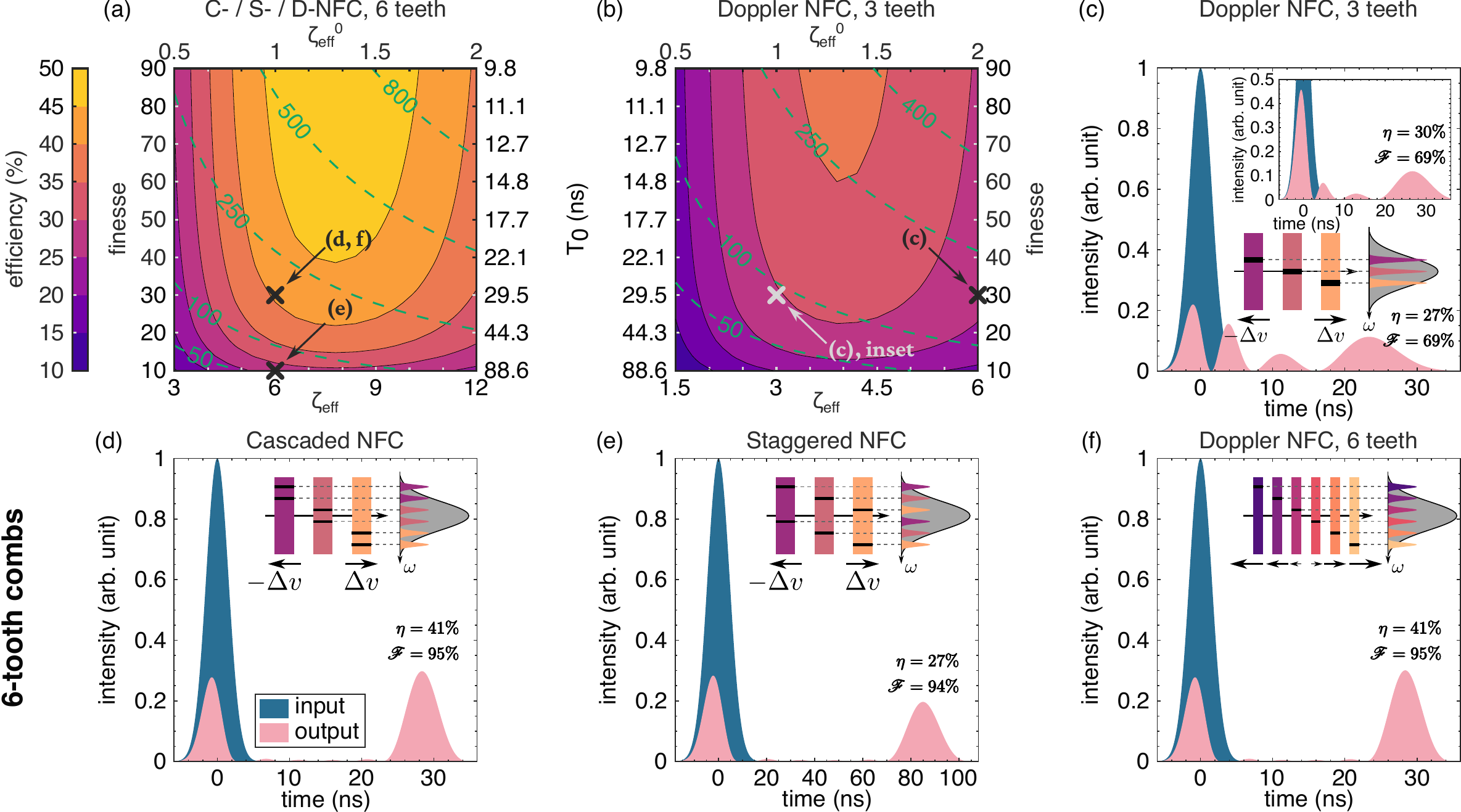}
    \caption{
    Hard X-ray quantum memory by static nuclear frequency combs.
    (a,b) QM efficiencies as functions of effective optical thickness $\xi_\text{eff}$ and comb finesse $\mathcal{F}$ for (a) 6-tooth C-NFC or S-NFC, and (b) 3-tooth D-NFC, all implemented with 3 M{\"o}ssbauer absorbers.
    The incident photon bandwidth is matched to (a) the C-NFC and/or S-NFC, which is twice that of (b) D-NFC for maintaining the same comb spacing $\Delta \omega$.
    Cyan dashed lines are contours of constant total resonant optical thickness, given by $\xi = N K \mathcal{F} \xi_\text{eff}^0$.
    (c-f) Demonstrations of QM using different static NFCs: (c) a 3-tooth D-NFC with 3 absorbers, (d) a 6-tooth C-NFC with 3 absorbers, (e) a 6-tooth S-NFC with 3 absorbers, and (f) a 6-tooth D-NFC composed of 6 absorbers.
    The parameter sets used in (c-f) are indicated by black and while crosses in (a) and (b).
    Panels (c) and (d) compare the performance of 3-tooth D-NFC and 6-tooth C-NFC, both implemented with 3 absorbers of equal total optical thickness $\xi=180$ and comb finesse $\mathcal{F}=30$, with the same pre-determined echo time $T_0=2\pi/(\mathcal{F}\Gamma)=29.53$~ns.
    Accordingly, the individual effective optical thickness is $\xi_\text{eff}^0=2$ for the 3-tooth D-NFC --- twice that of the C-NFC --- leading to strong reabsorption in (c). This reabsorption is mitigated in the inset by halving the optical thickness to achieve $\xi_\text{eff}^0=1$.
    Panels (d-f) compare three different 6-tooth NFCs, all with the same total optical thickness $\xi=180$. In (e), S-NFC finesse is set to one-third that of (d) C-NFC to preserve the same internal hyperfine magnetic field, resulting in a threefold increase in echo time and a modest drop in efficiency due to decoherence.
    Panel (f) shows that equivalent QM performance to (d) can be achieved using a 6-tooth D-NFC with the same $\xi=180$ and $\mathcal{F}=30$, but at the cost of doubling the number of absorbers.
    In all cases (a-f), the incident photon has a Gaussian temporal wave form with FWHM field duration of $\Delta t = 8\ln(2) / (5\Delta\omega + \Gamma)$.
    }
    \label{fig:static_NFC}
\end{figure*}

The incorporation of internal degenerate transitions into the NFC scheme enables a higher storage bandwidth for a given storage time $T_0$.
This is illustrated in Fig.~\ref{fig:static_NFC}~(a,~b), which compares the QM efficiencies of the 6-tooth C-NFC and the 3-tooth D-NFC.
For a given spectral spacing $\Delta \omega$, the former comb has a bandwidth on the order of $6 \Delta \omega$ while the latter is $3 \Delta \omega$.
The full-width at half-maximum (FWHM) field duration of the incident hard X-ray photon is adjusted
to spectrally matching the bandwidth of C-NFC but exceeding the bandwidth of D-NFC.
The magnetic-Doppler NFC is implemented using \ce{^{57}FeBO3}, which features a strong internal hyperfine field $\vec{B}_\text{hf}$ that lifts the nuclear spin degeneracy, while the D-NFC employs a non-magnetic medium such as \ce{^{57}Fe}-enriched stainless steel.
Both frequency combs utilize three \ce{^{57}Fe} M{\"o}ssbauer absorbers, maintaining the same level of mechanical complexity.
However, across storage times ranging from $10$~ns to $90$~ns, the 3-tooth D-NFC achieves a maximum efficiency of $\eta = 36\%$, whereas the 6-tooth C-NFC reaches up to $50\%$.

Figure~\ref{fig:static_NFC} (c, d) presents specific examples of QM using a 3-tooth D-NFC and a 6-tooth C-NFC, respectively, both implemented with the same total resonant optical thickness $\xi = 180$ and comb spacing $\Delta\omega = 30\Gamma$.
This corresponds to a velocity spacing of $\Delta v = 2.92$~mm~s\textsuperscript{-1} for the D-NFC, and combination of hyperfine field $B_\text{hf}=15.6$~T and velocity spacing $\Delta v = 5.83$~mm~s\textsuperscript{-1} for the C-NFC.
The QM efficiency $\eta$ and fidelity $\mathscr{F}$ (see the Supplementary Information~\cite{SI} for the definitions) are $\eta = \text{40.54\%}$, $\mathscr{F} = \text{94.83\%}$ for C-NFC, and $\eta = \text{27.22\%}$, $\mathscr{F} = \text{69.23\%}$ for D-NFC, respectively.
Since the individual effective optical thicknesses $\xi_\text{eff}^0$ differ due to the smaller number of comb teeth in the D-NFC, the inset of Fig.~\ref{fig:static_NFC}~(c) shows a 3-tooth D-NFC case adjusted to match the $\xi_\text{eff}^0 = 1$ used in Fig.~\ref{fig:static_NFC}~(d), resulting in $\eta = \text{29.56\%}$ and $\mathscr{F} = \text{69.04\%}$ --- both lower than those of the C-NFC.

Figure~\ref{fig:static_NFC} (d–f) compares three different configurations of the NFC: the C-NFC, the S-NFC, and the Doppler-only D-NFC, all with 6 spectral teeth and implemented by identical total effective resonant optical thickness $\xi_\text{eff} = 6$.
In each case, the incident photon remains bandwidth-matched with the respective comb.
The C-NFC features a frequency spacing of $\Delta \omega = \Delta \omega\Z = 30\Gamma$ and a Doppler frequency shift of $\Delta \omega\D = K \Delta \omega$, realized via a hyperfine magnetic field $B_\text{hf} = 15.6$~T combined with a velocity spacing $\Delta v = 5.83$~mm~s\textsuperscript{–1}, as discussed previously.
The S-NFC retains the same hyperfine field $B_\text{hf}$ as the C-NFC but moves the absorbers such that $\Delta \omega = \Delta \omega\D = \Delta \omega\Z / N = 10\Gamma$. This is implemented using a velocity spacing $\Delta v = 0.972$~mm~s\textsuperscript{–1}.
The 6-tooth D-NFC uses no hyperfine field and is constructed with a velocity spacing of $\Delta v = 2.92$~mm~s\textsuperscript{–1} to match the comb spacing $\Delta \omega = 30\Gamma$ for direct comparison with the C-NFC.
The results indicate that the C-NFC achieves QM performance comparable to the D-NFC, but with a $K$-fold reduction in mechanical complexity.
Meanwhile, the S-NFC enables hard X-ray photon echoes delayed by a factor of $N$ through a simple reduction of the velocity spacing by $N$-times --- a fast and precise adjustment that does not require modifying the hyperfine magnetic field amplitude via temperature control.

\section{On-demand quantum memory by time-reversal of the NFC dynamics}
\label{sec_on_demand_QM}

On-demand quantum memory can be implemented by full or partial time-reversal of the field evolution in a resonant medium in a controlled manner.
For the proposed magnetic-Doppler NFC, this entails simultaneously reversing the directions of both the hyperfine magnetic fields $\vec{B}_\text{hf}$ and the velocities $v_n$ of all absorbers, i.e., with $\vartheta\D = \vartheta\Z = 1$.
As discussed in Section~\ref{sec_DZNFC}, the former can be achieved within $\sim 4$~ns~\cite{Shvydko94Gerdau} by flipping the direction of the weak external magnetic field $\vec{B}_\text{ex}$, while the latter can be implemented by mounting the two outer absorbers on oppositely oriented, identical piezoelectric actuators driven by a common voltage signal.
Reversing these control parameters at time $T_\text{sw}$ inverts the phase evolution of the polarization waves, resulting in constructive interference at $2T_\text{sw}$ and producing an on-demand X-ray photon echo.
To minimize photon loss, the switching time should satisfy $T_\text{sw} < T_0$, allowing this scheme to support on-demand QM with storage times up to $2T_0$.

In the case of a Doppler-only frequency comb (D-NFC), this concept gives rise to the stepwise gradient echo (SGE) protocol for on-demand hard X-ray quantum memory~\cite{Zhang19Kocharovskaya}. However, synchronizing the control of multiple M{\"o}ssbauer absorbers remains experimentally challenging.
In contrast, the internal hyperfine magnetic fields $\vec{B}_\text{hf}$ in \ce{^{57}FeBO3} can be switched rapidly and reliably, without requiring precise amplitude control of the external magnetic field $\vec{B}_\text{ex}$.

\begin{figure}[hbt!]
    \centering
    \includegraphics[width=\linewidth]{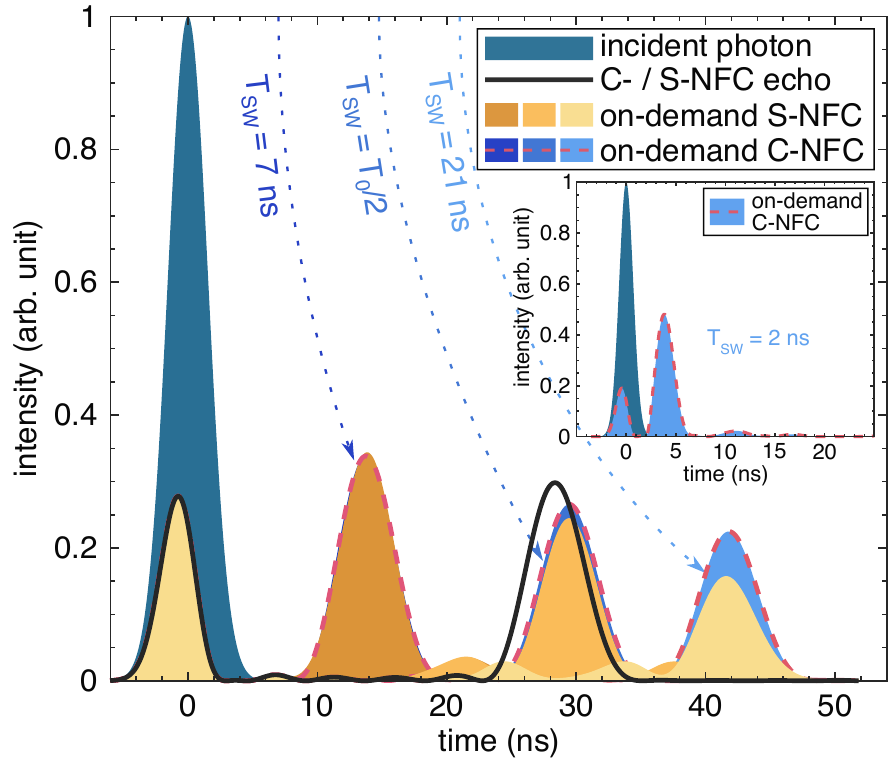}
    \caption{
    On-demand echo retrieval for C-NFC (pink dashed line with blue-filled area) and S-NFC (red-filled area) by varying $T_\text{sw}$ from 7~ns to 21~ns, with $\xi_\text{eff}=6$ and $\mathcal{F}=30$. The incident photon is illustrated in filled navy-blue area. The predetermined photon echo from static S-NFC or C-NFC (see the Supplementary Information~\cite{SI}) without performing any switch operation is plotted in black solid line for comparison.
    The inset shows an example of on-demand QM with an efficiency $57.65\%$ and fidelity $96.87\%$, using the C-NFC configuration with $\xi_\text{eff}=7.8$, $\mathcal{F}=66$, and a switching time of $T_\text{sw}=2$~ns.
    }
    \label{fig:FeBO3_various_Tsw_line_shape}
\end{figure}

Examples of on-demand QM using magnetic-Doppler NFCs are shown in Fig.~\ref{fig:FeBO3_various_Tsw_line_shape}. Both the C-NFC and S-NFC configurations are implemented using three \ce{^{57}FeBO3} absorbers, with the total resonant optical thickness fixed at $\xi = 180$, and comb spacing $\Delta \omega = 30\Gamma$.
The hyperfine magnetic fields and absorber velocities are simultaneously switched at time $T_\text{sw}$.
Without these switches, the efficiency of static NFCs is fundamentally limited to below $54\%$ due to photon reabsorption.
With active switching, however, the on-demand NFC configurations can surpass this limit, as demonstrated by the C-NFC retrieval of an X-ray photon with a storage time of $4$~ns and efficiency $58\%$ (Fig.~\ref{fig:FeBO3_various_Tsw_line_shape} inset). 
The capability for on-demand retrieval is further illustrated by varying $T_\text{sw}$ from $7$~ns to $21$~ns, resulting in X-ray photon echoes at the expected times $2T_\text{sw}$ with efficiencies of $45\%$, $38\%$, $34\%$ for C-NFC and $46\%$, $40\%$, $29\%$ for S-NFC, respectively.

\begin{figure}[hbt!]
    \centering
    \includegraphics[width=\linewidth]{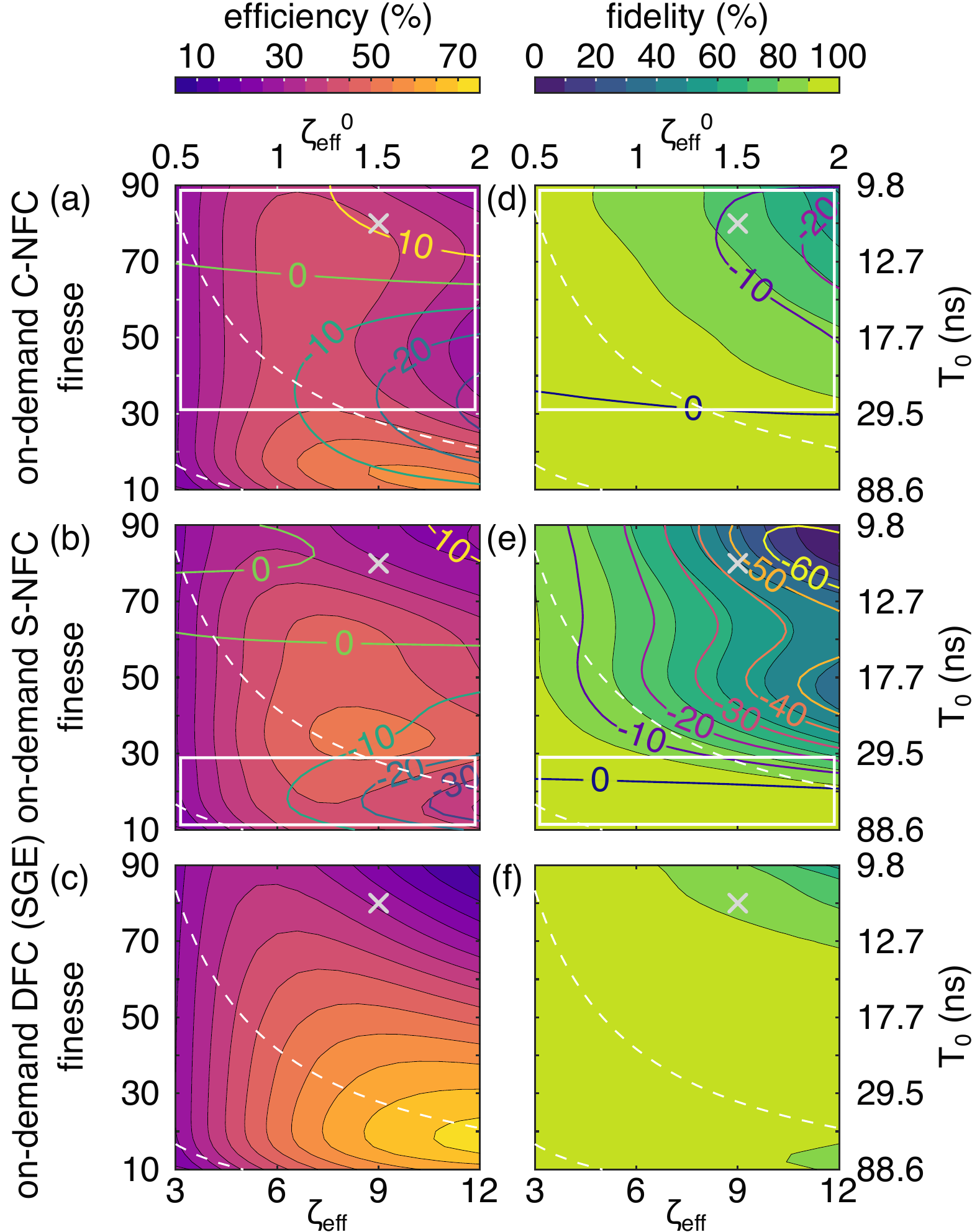}
    \caption{
    Performance of on-demand hard X-ray quantum memory using \ce{^{57}Fe}-based nuclear frequency combs.
    (a-c) Efficiencies and (d-f) fidelities of on-demand storage for $15$~ns of incident photons with Gaussian temporal wave form, characterized by a FWHM field duration $\Delta t = 8\ln(2) / (5\Delta\omega + \Gamma)$.
    Results are shown for: (a, d) on-demand C-NFC, (b, e) on-demand S-NFC, and (c, f) on-demand 6-tooth D-NFC (SGE).
    Numbered solid-color contour lines in panels (a, b) and (d, e) represent the efficiency and fidelity differences, respectively, relative to the on-demand D-NFC results shown in (c) and (f).
    White boxes enclose regions corresponding to feasible hyperfine magnetic fields in \ce{^{57}FeBO3}, while white dashed lines mark total optical thicknesses $\xi = 50$ and $\xi = 250$ across all panels.
    A visual illustration of on-demand retrievals in the regime of high total optical thickness $\xi$ and high comb finesse $\mathcal{F}$ --- where C-NFC and S-NFC exhibit higher efficiency but lower fidelity compared with D-NFC --- is provided in Fig.~S5 in the supplementary material, corresponding to the region marked by white crosses.
    }
    \label{fig:on_demand_NFC_contour}
\end{figure}

Figure~\ref{fig:on_demand_NFC_contour} compares the efficiencies and fidelities of on-demand QMs using C-NFC, S-NFC, and D-NFC (i.e., the SGE protocol), with all configurations employing six spectral teeth and a fixed switching time of $T_\text{sw} = 7.5$~ns. To facilitate comparison, the contour plots also show the relative efficiency and fidelity as percentages with respect to those of the SGE. 
The maximum achievable efficiencies $\eta_\text{max}$ for the C-NFC, S-NFC, and SGE schemes are found to be $56\%$, $52\%$, and $72\%$, respectively. 
Unlike in the static case, the on-demand C-NFC and S-NFC configurations exhibit notably different performance, both with lower $\eta_\text{max}$ than the SGE protocol.
Nevertheless, they maintain comparable overall efficiency and surpass the SGE protocol in the regime of high optical thickness and $\mathcal{F} \gtrsim 70$. 
However, the fidelities of both magnetic-Doppler configurations decrease more rapidly with increasing $\mathcal{F}$ and effective optical thickness $\xi_\text{eff}$, especially for S-NFC.
On the other hand, given the limited range of hyperfine magnetic fields available in \ce{^{57}FeBO3} and the practical constraint of a total optical thickness $\lesssim 200$, both on-demand C-NFC and S-NFC demonstrate lower efficiencies compared with on-demand D-NFC.
However, as discussed in Section~\ref{sec_DZNFC}, a key advantage of the magnetic-Doppler approach lies in its minimal --- or even absent --- requirement for mechanical synchronization.
Taking into account the trade-offs among storage efficiency, fidelity, and experimental feasibility, the on-demand C-NFC offers a practical and efficient route to hard X-ray quantum memory.

It is worth noting that a viable on-demand QM requires simultaneous reversal of both the hyperfine magnetic fields and the absorber velocities. Partial reversal --- switching only the magnetic field or only the velocities --- does not guarantee successful on-demand retrieval (see the Supplementary Information for details~\cite{SI}).

\section{Conclusion}
\label{sec_conclusion}

We propose a magnetic hyperfine-Doppler nuclear frequency comb for implementing quantum memory of hard X-ray photons. The comb is realized using three \ce{^{57}FeBO3} absorbers, where the intrinsic hyperfine magnetic field lifts the nuclear spin degeneracy, giving rise to non-degenerate hard X-ray transitions. 
This effectively doubles the number of comb teeth without increasing the number of absorbers.
Quantum memory with a predetermined storage time can be achieved using static frequency combs, where constructive interference among all frequency components of the polarization wave results in coherent re-emission after a fixed delay. 
In contrast, on-demand quantum memory is enabled by simultaneously reversing the hyperfine magnetic field --- through reversal of an external control field --- and the absorber velocities, thereby inducing a time-reversed phase evolution of the polarization waves and allowing retrieval at a controllable time.


Compared with the pure Doppler frequency comb, the magnetic–Doppler approach substantially reduces the need for complex mechanical synchronization required for on-demand retrieval.
Under feasible experimental conditions, and neglecting off-resonant losses, this scheme yields on-demand retrieval efficiencies of $45\%$, $38\%$, and $34\%$ for storage times of $14$~ns, $30$~ns, and $42$~ns respectively.

Due to the low ground-state spin multiplicity of \ce{^{57}Fe}, each absorber supports only two non-degenerate transition pairs, limiting the total number of comb teeth achievable with a given level of synchronous mechanical control.
However, the concept of a magnetic-Doppler nuclear frequency comb is not restricted to \ce{^{57}Fe}. It can be extended to other nuclei, such as \ce{^{181}Ta}, which features a $6.2$~keV transition energy with a long excited state lifetime of $8.73$~$\mu$s, and supports eight non-degenerate transitions per absorber.
Moreover, with adjancent differential Zeeman splitting reaching $\sim 207$ natural linewidths per Tesla, such a scheme can be implemented using external magnetic fields alone, without resorting to crystal environments with strong internal hyperfine fields.
In summary, the magnetic-Doppler nuclear frequency comb presents a practical and versatile approach for realizing on-demand quantum memory in the hard X-ray regime.

\section*{Acknowledgements}
The authors appreciate the financial support from the National Science Foundation (NSF, Grant No. PHY-240-97-34). Portions of this research were conducted with the advanced computing resources provided by Texas A\&M High Performance Research Computing. The authors also gratefully acknowledge insightful discussions with J{\"o}rg Evers and Ralf R{\"o}hlsberger.


\bibliographystyle{apsrev4-1}
\bibliography{BibTex_QM_FeBO3}

\end{document}